\documentclass[12pt]{article}%
\usepackage{amsfonts}
\usepackage{amsmath}%
\usepackage{amssymb}%
\usepackage{graphicx}
\providecommand{\U}[1]{\protect\rule{.1in}{.1in}}

\begin{document}

\title{Some remarks on an old problem of radiation and gravity}
\author{C. S. Unnikrishnan$^{1}$ and George T. Gillies$^{2}$\\$^{1}$\textit{Gravitation Group, Tata Institute of Fundamental Research, }\\\textit{\ Homi Bhabha Road, Mumbai - 400 005, India}\\$^{2}$\textit{School of Engineering and Applied Science,}\\\textit{University of Virginia}, \textit{Charlottesville, VA 22904-4746, USA}\\E-mail address: $^{1}$unni@tifr.res.in, $^{2}$gtg@virginia.edu\medskip\\Honorable mention in the Gravity Research Foundation 2014 \\Awards for Essays on Gravitation.}
\date{}
\maketitle

\begin{abstract}
The assumed universality of the equivalence principle suggests that a particle
in a gravitational field has identical physics to one in an accelerated frame.
Yet, energy considerations prohibit radiation from a static particle in a
gravitational field while the accelerating counterpart emits. Solutions to the
fundamental problems of radiation from charges in a gravitational field and
consequences to the equivalence principle usually contrast the far-field and
global nature of radiation with the local validity of the equivalence
principle. Here, we suggest reliable physical solutions that recognizes the
essential need for motional currents and the magnetic component for radiation
to occur. Our discussion reiterates the need for a fresh careful look at
universality of free fall for charged particles in a gravitational field.

\pagebreak

\end{abstract}%

The principle of equivalence (PE) is the firm and tested pillar on which the
elegant theory of gravitation -- the general theory of relativity -- is
beautifully constructed. The basic form of PE, called the Weak Equivalence
Principle (WEP), asserts that the local physical effects in a uniform
gravitational field are identical to those in a uniformly accelerated frame.
Extension of the WEP to all physics including gravitational phenomena is
called the Einstein Equivalence Principle (EEP), and it is assumed to be
valid. One established result in electrodynamics is that a uniformly
accelerated charge radiates. Yet, such a fundamental and important result
becomes a source of continued controversy and discussion when treated in the
context of the WEP and gravity \cite{review}. Several questions arise, for
which convincing concise answers remain somewhat elusive, even though
defendable answers exist as a result of prolonged debates. Simplest of these
questions is perhaps whether a charged particle that is freely falling in a
uniform gravitational field radiates. That it should radiate, if a uniformly
accelerated particle in standard electrodynamics radiates, seems certain but
the fact that there is no radiation reaction in uniform acceleration casts
doubts on an immediate assertion. Also, the particle is inertial for a
co-falling observer, who should not detect radiation if strict WEP is valid.
The standard solution is to say that while there is radiation in the far field
\ region in the frame of any inertial observer, the observer comoving with the
particle is in the near field region and will not detect any radiation,
consistent with WEP. In fact, this has been established in a careful paper by
Rohrlich \cite{Rohr1}. While many authors consider this as satisfactory
\cite{Boul}, there are some who rejected such a solution, including Feynman.
In his lectures in gravitation he argued \cite{Feyn} that a uniformly
accelerated charge would not radiate at all! He writes, "The Principle of
Equivalence postulates that an acceleration shall be indistinguishable from
gravity by any experiment whatsoever. In particular, it cannot be
distinguished by observing electromagnetic radiation. There is evidently some
trouble here, since we have inherited a prejudice that an accelerating charge
should radiate, whereas we do not expect a charge lying in a gravitational
field to radiate." \ Then he goes on to say that the standard equation for
power radiated by the charge
\begin{equation}
\frac{dW}{dt}=\frac{2e^{2}}{3c^{3}}a^{2}%
\end{equation}
is an approximation valid for only bounded or periodic motion and has "led us astray."%

\begin{figure}
[ptb]
\begin{center}
\includegraphics[
natheight=2.972400in,
natwidth=3.262100in,
height=1.8689in,
width=2.0505in
]%
{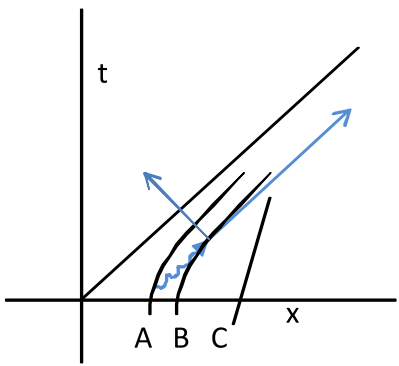}%
\caption{The necessity of absolute acceleration for radiation to occur can be
established even in the absence of gravity. Accelerated observer B at
\emph{relative rest with }\ an accelerating charge A should detect radiation
in the far zone, $d\gg\lambda\approx c^{2}/a.$ Also, a comoving observer along
A will detect `own' radiation after reflections, even though the charge is at
rest locally. No radiation should be detected by B from the inertial charge C
\ even though there is relative acceleration. }%
\end{center}
\end{figure}

Feynman's rejection of the standard expectation is reasonable considering that
even a comoving observer can eventually detect such radiation, if it is
emitted once, by reflection from any other observer far away (see fig. 1).
This indicates that the WEP has to be carefully formulated when the
possibility of radiation exists and the standard statement of complete
equivalence between a uniform gravitational field and an accelerating frame is
not adequate. In other words, one cannot solve such problems by appealing to a
near-world -- far-world separation because the two can be connected by the
retarded radiation. Another way of posing the problem is to ask what would be
the result of Galileo's experiments with freely falling objects if one of them
is electrically charged? Does it radiate and slow down a bit due to radiation
reaction, or does it fall with the same acceleration as a neutral object?
Fulton and Rohrlich \cite{Rohr-Fult} write "If Galileo had dropped a neutron
and a proton from the leaning tower of Pisa they would have fallen equally
fast...A charged and a neutral particle in a homogeneous gravitational field
behave exactly alike, except for the emission of radiation from the charged particle."

Now we present a simple argument to show that a Galileo experiment with a
charged and a neutral particle has to violate universality of free fall (UFF)
if standard electrodynamics is correct. The radiated power is $2e^{2}%
a^{2}/3c^{3}.$ The total energy of the particle is $E=mv^{2}/2-mgh.$ With
$\frac{dh}{dt}=v$ we have the conservation requirement
\begin{equation}
\frac{2e^{2}a^{2}}{3c^{3}}=-\frac{dE}{dt}=m\vec{g}\cdot\vec{v}-m\vec{v}%
\cdot\vec{a}=m\vec{v}\cdot(\vec{g}-\vec{a})
\end{equation}
Therefore, even the slightest of radiation implies $\left(  \vec{g}-\vec
{a}\right)  \neq0,$ violating UFF. The importance of settling this one way or
other experimentally cannot be overstated.

Assuming that it is established that a uniformly accelerated charged particle
in flat space radiates, the equivalence principle suggests that a charged
particle at rest in a locally uniform gravitational field should emit
radiation. This surprising consequence of the naive application of the WEP has
attracted much attention, and much has been written about it \cite{Ginzberg95}
Most analysis starts with the expectation that such a static charge should
not radiate and reasons are invented for the absence of radiation. However,
there are some exceptions to this claimed by those who contend that a static
charged particle in a gravitational field indeed should emit radiation
\cite{Harpaz}. The most accepted argument against radiation by a static charge
in a gravitational field is the impossibility of finding a gravitational field
of infinite extent that would exactly correspond to true hyperbolic, uniformly
accelerated motion of a charged particle \cite{Bondi}. However, this global
and teleological argument is clearly not satisfactory because the actual
result of radiation in electrodynamics is arrived at from a part of the
accelerated trajectory and nowhere in the electrodynamic theory does one
assume that the entire trajectory has to be assured to be infinite and
hyperbolic before the charge decides to emit radiation! Therefore, there is
compelling motivation to find a more satisfactory and local reason why a
static charge in a gravitational field does not radiate.

We present two reliable arguments to show that a static charge in a
gravitational field cannot radiate. Since the energy that is radiated is
always positive, proportional to the square of the radiation field amplitude,
this has to come from the bound gravitational energy of the charged particle
by moving to a more negative gravitational potential, as given by eq. 2. Since
$v=0,$ $\frac{dE}{dt}=0$ \ and hence, the stationary charge in the
gravitational field cannot radiate. In fact, the usual expectation that the
static charge in a gravitational field does not radiate derives from such a
requirement of energy conservation.

Another of our general argument relies on the characteristic feature of
electromagnetic radiation that \emph{both} electric and magnetic fields,
coupled to each other through their time dependence, are essential for
radiation to occur. Explicitly,
\begin{equation}
\frac{\partial\vec{B}}{\partial t}=-\nabla\times\vec{E},\quad\mu
_{0}\varepsilon_{0}\frac{\partial\vec{E}}{\partial t}+\mu_{0}j=\nabla
\times\vec{B}%
\end{equation}

However, both empirically and theoretically, time dependent fields can be
generated only by time dependent currents somewhere in space. Charge currents
are essential for creating vector potentials and magnetic fields. The crucial
difference between the field of a charged particle in a gravitational field
and one that is uniformly accelerating is the absence of a magnetic component
in the former situation. Though the electric fields may have identical spatial
curvatures as a function of distance from the charge in both cases, a magnetic
component is present only in the case of an accelerating charged particle. It
is obvious that an electric field does not manifest as a magnetic field in the
presence of a gravitational field whereas it does when there is motion. This
observation immediately solves the fundamental problem we started with - does
a static charge in a gravitational field radiate? If it doesn't, why not? \ We
have the physically reliable answer that it does not radiate, because static
gravity does not generate the magnetic component that is essential for
electromagnetic radiation.

We know from elementary considerations analogous to that in electrodynamics
that a massive particle that is accelerating radiates gravitational waves.
Constraints from momentum conservation and the feature that there is only type
of gravitational charge (mass) implies that this radiation is quadrupolar
instead of dipolar, however. Given that there is radiation from accelerated
masses, important fundamental issues arise in the context of EEP. Does gravitational
radiation from a freely falling particle conflict with UFF and WEP? Does a
stationary massive particle in a gravitational field radiate?

A massive particle that is freely falling in a gravitational field should emit
gravitational radiation because it is the source of time dependent `magnetic'
and `electric' metric components $h_{0i}$ and $h_{00}$. Indeed, this solution
is suggested by the approximated (vacuum) Einstein's equations in terms of
these components of the gravitational field \cite{GEM},%

\begin{equation}
\frac{\partial\vec{B}_{g}/2}{c\partial t}=-\nabla\times\vec{g},\quad\frac
{1}{c}\frac{\partial\vec{g}}{\partial t}=\nabla\times\vec{B}_{g}%
/2,\ \text{with }\vec{B}_{g}=\nabla\times\vec{A},\;A_{i}/c^{2}=-h_{0i}/2
\end{equation}
It is interesting and important to note that unlike the situation for
electromagnetic radiation where UFF would be violated due to the
non-universality of the $e/m$ ratio, the gravitational case will respect UFF
in spite of escaping radiation because $m_{g}/m_{i}$ is universal. This is a
subtle point on close examination because the dynamics of `back-action', in
spite of the name given to the phenomenon, is controlled by energy loss rather
than a force acting back on the particle. Nevertheless, the back-action is
proportional to the inertial mass while the energy loss is proportional to the
gravitational mass and their equivalence ensures UFF.

Energy considerations alone suggests that a static charge in a gravitational
field should not radiate. However, for gravitational problems where the global
aspects of associated space-times might enter the discussion, such answers may
not give the complete picture. Also, if this is indeed the answer, one is at a
loss to explain the apparent inconsistency with the EEP. Here again, our
analysis provides the general framework for a clear answer. Gravitational
radiation also requires a `magnetic' component, in the form of
gravito-magnetism linked to relative motion or current of the massive
particles and the metric components $g_{0i}$. Since the static mass in a
gravitational field is not a source of gravito-magnetic fields, radiation
cannot occur.

The problem and our solution have larger implications when we transfer the
scenario to quantum vacuum in a gravitational field and in an accelerated
frame. It is well known, as the Unruh effect \cite{Unruh}, that uniformly
accelerated observers with an apparent horizon should see the vacuum
transformed into a thermal background with temperature proportional to the
acceleration. Similar results about the effective temperature proportional to
the surface gravity of a blackhole, and the possibility of Hawking evaporation
are also well studied and discussed \cite{Hawk-review}. However, whether any
gravitating body, like a dense neutron star, and not just those with a
horizon, convert vacuum to thermal background has never been satisfactorily
answered. Based on our analysis that establishes a broken equivalence between
a static gravitational field and an accelerated frame when radiation problems
are considered, we may assert, at least as a robust conjecture, that an
observer in a static gravitational field will not see the vacuum transformed
into a thermal radiation background.

We conclude that there are physical phenomena where a uniform gravitational
field is not locally equivalent to an accelerating frame. However, this does
not violate the WEP because the inequivalence can be traced to the radiative
components in the global sum.

\end{document}